\documentclass[aps,prl,twocolumn,floatfix,citeautoscript,superscriptaddress]{revtex4-1}

\usepackage{amsmath}
\usepackage{amssymb}
\usepackage{amsfonts}
\usepackage{graphicx,graphics}
\usepackage{bm}
\usepackage{multirow}

\begin{document}

% \title{Ultra-high mobility in flux-grown monolayer WSe$_2$}
\title{Transport Study of Charge Carrier Scattering in Monolayer WSe$_2$}

\author{Andrew Y. Joe}
% \email{Equal contribution}
\affiliation{Department of Physics, Harvard University, Cambridge, MA 02138, USA}
\affiliation{Department of Physics and Astronomy, University of California, Riverside, CA 92521, USA}
\author{Kateryna Pistunova}
% \email{Equal contribution}
\affiliation{Department of Physics, Harvard University, Cambridge, MA 02138, USA} 
\author{Kristen Kaasbjerg}
% \email{kkaa@dtu.dk}
\affiliation{Center for Nanostructured Graphene (CNG), Department of Physics, Technical University of Denmark, DK-2800 Kongens Lyngby, Denmark}
\author{Ke Wang}
\affiliation{Department of Physics, Harvard University, Cambridge, MA 02138, USA}  
\author{Bumho Kim}
\affiliation{Department of Mechanical Engineering, Columbia University, New York, NY 10027, USA}
\author{Daniel A. Rhodes}
\affiliation{Department of Mechanical Engineering, Columbia University, New York, NY 10027, USA}
% \author{Bumho Kim}
% \affiliation{Department of Mechanical Engineering, Columbia University, New York, NY 10027, USA}
\author{Takashi Taniguchi}
\affiliation{National Institute for Materials Science, Tsukuba, Japan}
\author{Kenji Watanabe}
\affiliation{National Institute for Materials Science, Tsukuba, Japan}
\author{James Hone}
\affiliation{Department of Mechanical Engineering,
  Columbia University, New York, New York 10027, USA}
\author{Tony Low}
\affiliation{Department of Electrical and Computer Engineering, University of Minnesota, Minneapolis, MN 55455, USA}
\author{Luis A. Jauregui}
\affiliation{Dept. of Physics and Astronomy, The University of California, Irvine, California
  92697, USA} 
\author{Philip Kim}
\email{philipkim@g.harvard.edu}
\affiliation{Department of Physics, Harvard University, Cambridge, MA 02138, USA} 
\date{\today}

\begin{abstract}
    Employing flux-grown single crystal WSe$_2$, we report charge carrier scattering behaviors measured in $h$-BN encapsulated monolayer field effect transistors. We perform quantum transport measurements across various hole densities and temperatures and observe a non-monotonic change of transport mobility $\mu$ as a function of hole density in the degenerately doped sample. This unusual behavior can be explained by energy dependent scattering amplitude of strong defects calculated using the T-matrix approximation. Utilizing long mean-free path ($>$500 nm), we demonstrate the high quality of our electronic devices by showing quantized conductance steps from an electrostatically-defined quantum point contact. Our results show the potential for creating ultra-high quality quantum optoelectronic devices based on atomically thin semiconductors.
\end{abstract}
\maketitle

%\section{Introduction}

Two-dimensional (2D) monolayers of transition metal dichalcogenides (TMDs;
$MX_2$) hold great promise for future electronics and
optoelectronics~\cite{Heinz:ThinMoS2,Kis:MoS2Transistor,Schuller:Photocarrier,avouris20172d}. Due
to their strong spin-valley coupling~\cite{Yao:SpinValley,Heinz:Spin}, they are
potential candidates for spin- and valleytronics applications for which
high-mobility samples with long spin and valley lifetimes are essential. Progress towards using TMDs for engineering applications or exploring strongly-correlated Quantum-hall states have been hindered by low carrier mobilities in comparison to other 2D electron gas (2DEG) systems such as graphene or GaAs \cite{Dean2010, Chung2021}. Similar to conventional 2D semiconductor heterostructure systems, experimental low-temperature mobilities in monolayer TMDs are most often limited by
short-range and Coulomb disorder
scattering~\cite{Kis:Engineering,Herrero:Intrinsic,Avouris:Electronic,Wang:Towards,Eda:Transport,Eda:Charge,Hone:Multi,Eda:Quantum,Hone:Low},
and have only recently reached values exceeding
1000~cm$^{-2}$/(V$\cdot$s)~\cite{Tutuc:Contacts,Ensslin:Gate,Gustafsson2018,Tutuc:Large,Ensslin:Interactions}.

%Theory:
%\begin{itemize}
%\item charged impurity
%  scattering~\cite{Fischetti:Mobility,Jena:ChargeScattering}.
%\item point defects~\cite{Heine:Defect,Guinea:Effect,Sousa:Anomalous}.
%\item Quantum transport
%studies~\cite{Shen:Intervalley,Xiao:Spin,Guinea:Quantum,Falko:SpinValley,Peeters:Quantum,Burkard:Landau,Sousa:%Anomalous,Houzet:Weak,Jauho:Electron}.
%\end{itemize}

%Exp. transport studies:
%\begin{itemize}
%\item thermally activated hopping
%  transport~\cite{Ghosh:Nature,Wang:Hopping,Wang:Towards,Wang:Probing,Morpurgo:Hole}.
%\item \cite{Ghosh:Nature,Herrero:Intrinsic,Kis:Engineering,Eda:Transport,Wang:Towards,Wang:High,Wang %:Realization} 
%\item Mobility, magneto and quantum transport in 2D
%  TMDs~\cite{Kis:Electrical,Hone:Multi,Eda:Quantum,McEuen:The,Tutuc:Shubnikov,Hone:Low,Ensslin:Gate,Dean:Ambi,%Tutuc:Large,Ensslin:Interactions}.
%\item Disorder scattering by atomic vacancies is often mentioned as a
%  mobility-limiting
%  factor~\cite{Wang:Towards,Eda:Charge,Eda:Transport,Hone:Multi,Eda:Quantum}
%\item Low-T mobility $\mu=e\tau_\text{tr}/m^*$ probes directly the transport relaxation time $\tau_\text{tr}$ and unique information about the dominating type of disorder scattering can be inferred by studying the density dependence of $\mu$. 
%\end{itemize}

Vast improvements in the quality of TMD materials have been made utilizing a flux growth technique, decreasing the density of point defects in WSe$_2$ from $10^{13}$ cm$^{-2}$ to below $10^{11}$ cm$^{-2}$\cite{Edelberg:2019,Liu:2023}. Initial transport measurements in these samples have reported mobilities reaching 840 cm$^2$/(V$\cdot$s) at room temperature and exceeding 44,000 cm$^2$/(V$\cdot$s) at low temperatures~\cite{Liu:2023}. These improvements over crystals grown by the previously conventional chemical vapor transport (CVT) method have shown to be critical for realizing strongly correlated physics in TMD monolayers and heterostructures \cite{Gustafsson2018,Shi2020,Wang2020,Shi2022}. However, careful transport studies of the nature of these defects and also the comparison of CVT and flux-grown crystal devices have not been performed.

In this work, we demonstrate unprecedented transport properties in archetypal monolayer WSe$_2$ based devices fabricated with CVT and flux growth crystals showing ultrahigh mobilities. We measure hole mobilities as large as 25,000 cm$^2$/(V$\cdot$s) for flux-grown samples at low temperatures, whereas we are limited to $\sim$ 3000 cm$^2$/(V$\cdot$s) in the CVT samples. Interestingly, the mobility ($\mu$) in both CVT and flux-grown crystals shows an unconventional dependence on the carrier density ($n$) which increases at low $n$ while decreasing at high $n$.
The non-monotonic behavior of the mobility with $n$, not expected for long ranged Coulomb scatterers~\cite{Hwang:Universal}, suggests that transport is dominated by screened short range scatterers at high densities~\cite{Hwang:Short}.

%The latter is unexpected for single-band (valley degenerate) transport in 2D semiconductor systems \cite{Hwang:Short,Hwang:Universal} where we anticipate a $\tau^{-1}\sim n^{-\alpha}$ (with $\alpha>0$) scaling for the transport relaxation time, implying a $\mu\sim n^\alpha$ scaling for the mobility.

By inspecting the transport and quasiparticle (quantum) scattering times obtained from the measured Hall mobility and SdH oscillations, we show that the transport characteristics are consistent with a nontrivial interplay between (i) disorder scattering due to \emph{intrinsic} atomic defects, such as, e.g., commonly encountered atomic vacancies, and (ii) scattering by \emph{extrinsic} (remote) charge impurities in the substrate. This manifests itself in the peculiar situation where, at low temperatures, the transport lifetime is limited by the former, while the quantum lifetime is limited by the latter (new paradigm).

\begin{figure}[b]
  \centering
  \hspace{1mm}
  \includegraphics[width=0.99\linewidth]{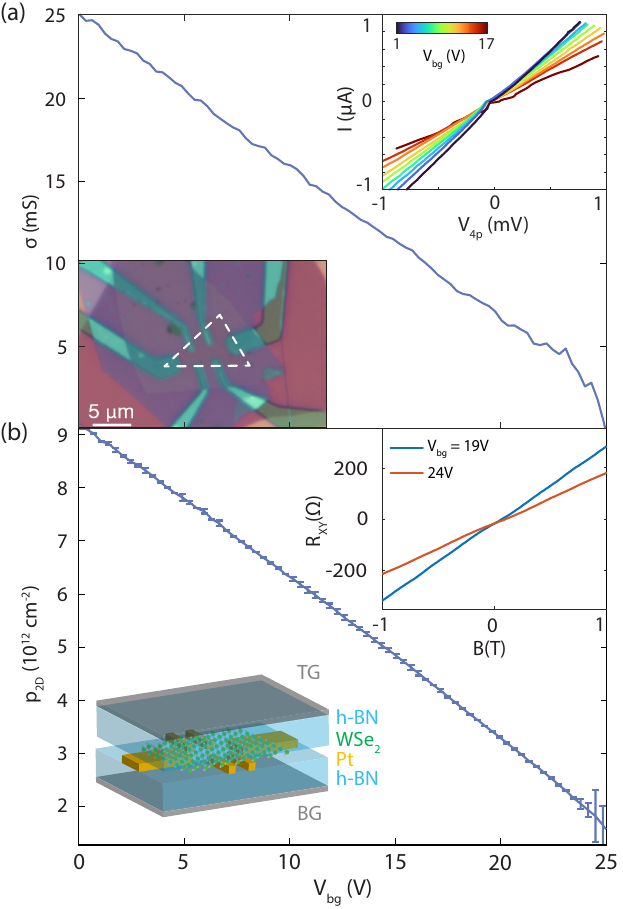}
  \caption{(a) Conductivity $\sigma$ measured at $T=1.5 K$ with back gate voltage $V_{BG}$.  Upper inset: Four probe IV curves taken at $T=1.5$~K at different values of $V_{BG}$ while applying a large negative $V_{TG}$ to dope the contact area. The linearity of IV curves proves ohmic p-type contact to single layer WSe$_2$ at cryogenic temperatures. Lower inset: Optical image of a representative device with top and bottom graphite gates and pre-patterned Pt contacts. WSe$_2$  monolayer is outlined in white dashed line. (b) Hole density obtained from Hall measurements, with $V_{BG}$. Upper inset: Representative transverse resistance $R_{xy}$ curves at different $V_{BG}$ with magnetic field $B$. Lower Inset: cross section cartoon of a representative device.}
\label{fig:1}
\end{figure}

We fabricate dual graphite gated, $h$-BN encapsulated single layer WSe$_2$ devices with pre-patterned platinum (Pt) contacts \cite{Tutuc:Contacts,Jauregui:OurPaper}. We transfer the monolayer WSe$_2$ on top of the pre-patterned Pt contacts, which has a high work function that matches the valence band edge of WSe$_2$ (see SI for contact engineering details). We use thin, optically transparent graphite gates to allow optical access to the WSe$_2$ flake. The thickness of top and bottom hBN gate dielectric is 50 and 73~nm, respectively. The lower inset of Fig. \ref{fig:1}a and Fig. \ref{fig:1}b shows an optical image and cross-sectional diagram of a representative WSe$_2$ device made with mechanically exfoliated flux-grown crystals.

To activate the contacts, we apply a top gate voltage of V$_{tg} = -23.1$ V, doping the TMD flake in the channel and in the contact area. Since the bottom gate is screened by the pre-patterned electrodes, we can tune the channel carrier density using the bottom gate while maintaining high contact doping. The upper inset of Fig. 1a shows linear four-probe ($V_{4p}$) IV curves at different channel carrier densities at 1.5 K, demonstrating Ohmic p-type contacts to monolayer WSe$_2$ at cryogenic temperatures. By highly doping the contact area and using pre-patterned Pt contacts, we achieve a low barrier for injecting holes. Ohmic contacts to single layer WSe$_2$ allow us to study its magnetotransport. To extract transport mobility, we measure conductivity at different bottom gate voltages ($V_{BG}$), as shown in Fig. \ref{fig:1}a. In Fig. \ref{fig:1}b, we obtain the hole density ($p_{2D}$) from Hall measurements at different gate voltages. The linearity of the transverse resistance $R_{xy}$ curves further proves the quality of the electrical contact and gives us a lower limit for our tunable doping of $p_{2D} \approx 1.5\times 10^{12}$ cm$^{-2}$.

\begin{figure}[!t]
  \centering
  \includegraphics[width=0.99\linewidth]{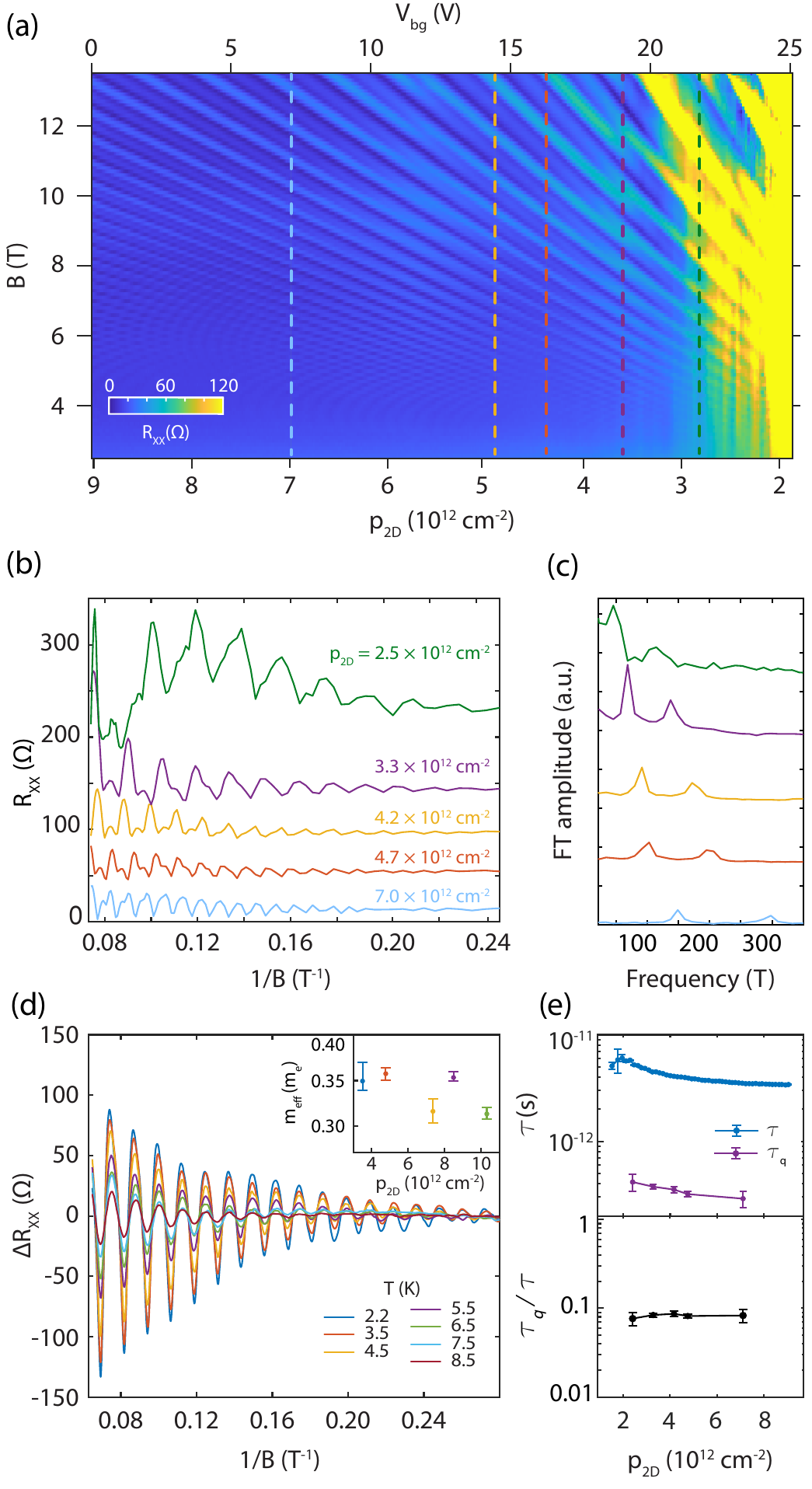}
  \caption{(a) Landau fan diagram of longitudinal resistance $R_{xx}$ measured at $T=1.5$ K as a function of bottom gate $V_{BG}$ voltage and magnetic field. Dashed lines correspond to line cuts in (b). (b) $R_{xx}$ for different bottom gate voltages as a function of the inverse magnetic field (1/$B$). (c) Fourier transform amplitude with frequency, spaced proportionally to back gate voltage. Colors correspond to the data in (b). (d) Change in longitudinal resistance $\Delta R_{xx}$ of a CVT device as a function of the inverse of magnetic field measured at various temperatures. Inset: effective hole mass with hole density. (e) (Top) Transport scattering lifetime, $\tau$, and quantum scattering lifetime, $\tau_q$, as a function of hole density. (Bottom) Ratio $\tau_q/\tau$ of quantum scattering lifetime to transport scattering lifetime as a function of hole density.
}
\label{fig:2}
\end{figure}

Under a perpendicular magnetic field ($B$) our samples exhibit Shubnikov–de Haas (SdH) oscillations. Figure \ref{fig:2}a shows a Landau fan diagram with longitudinal resistance $R_{xx}$ measured at $T=1.5$~K as a function of bottom gate $V_{BG}$ in fields up to 13.5~T. The switching between even and odd integer denominator Landau Level (LL) observed in the fan diagram, consistent with previous studies, arises due to the interplay between the Zeeman splitting and the cyclotron energy \cite{Tutuc:Contacts,Gustafsson2018}. If the Zeeman splitting is equal to or larger than the cyclotron energy, the LL sequence changes depending on the $E_{Z}/E_{c}$ ratio. A ratio close to an even (odd) integer leads to a sequence that is dominated by odd (even) states. Fig. \ref{fig:2}b show line cuts of the longitudinal resistance $R_{xx}$ vs inverse magnetic field at $V_{BG}$ voltages corresponding to the dashed lines in Fig. \ref{fig:2}a. Fig. \ref{fig:2}c shows the corresponding Fourier transform (FT) amplitude vs frequency, showing the principle frequency ($f$) and its second harmonic (2$f$), revealing the even-odd effect mentioned above. We calculate the SdH density $p_{SdH}=(2e/h)f$, which agrees well with a LL degeneracy of 2 (i.e., $p_{2D}=2p_{SdH}$). Figure \ref{fig:2}d shows $\Delta R_{xx}$ vs $1/B$ from the CVT device (see SI) at different temperatures, which we fit to the Dingle factor to extract a hole effective mass $m^* = 0.35m_{0}$, close to the value obtained in a previous study~\cite{Tutuc:Contacts}. 

From the Dingle analysis of SdH oscillation amplitude vs $1/B$, we extract the quantum scattering lifetime $\tau_q$ and compare it with the transport scattering lifetime $\tau$, estimated from $\tau=\sigma m^*/e^2 p_{2D}$, as function of hole carrier density $p_{2D}$ (Fig. 2(e)). We find that both $\tau$ and $\tau_q$ increase with decreasing density for $p_{2D}>p_*=$2$\times$10$^{12}$cm$^{-2}$, reaching the maximum values, $\sim$6000 fs and $\sim$600 fs, respectively. Below $p_*$, $\tau$ decreases steeply as $p_{2D}$ decreases further, while $\tau_q$ cannot be estimated in this regime due to disappearance of the SdH oscillation. 

The observed behavior of $\tau$ and $\tau_q$ suggests that there is an intricate interplay between the short-range and long-range scatterers in our samples. First, the decreasing $\tau$ with decreasing $p_{2D}<p_*$ suggests that the long ranged charged Coulomb scatters dominates in this lower density regime as the carrier screening becomes weaker. Similar behaviors of $\tau$ in the low density limit was obtained in the commercially obtained CVT grown crystals (see SI), although $p_*$ in this lower quality sample tends to be much higher than that of the flux grown samples. Since both the flux grown and the CVT samples were measured in similar device geometry, we speculate the unscreened long-range scatterers are likely extrinsic to the WSe$_2$ channels. Second, decreasing $\tau$ and $\tau_q$ in the higher density regime, $p_{2D}>p_*$, suggests that strong energy dependent scattering rate for short-range scatterers. We note that in this high density limit, the long-range Coulomb scatters are screened, leaving effect of short-range scatterers only. From a simple Born approximation based on a weak first order perturbation~\cite{Hwang:Short}, one expects that $\tau \sim \tau_q$ that can be remain density independent. However, our experimental observation clearly indicates that both $\tau$ and $\tau_q$ decreases with increasing $p_{2D}$, while $\tau_q/\tau \sim$ 0.1. This unusual trends of scattering times, thus, suggest that one needs to consider the effect of strong short range scatterers beyond the Born approximation~\cite{Kaasbjerg2020}.  

\begin{figure}[!t]
  \centering
  \includegraphics[width=0.99\linewidth]{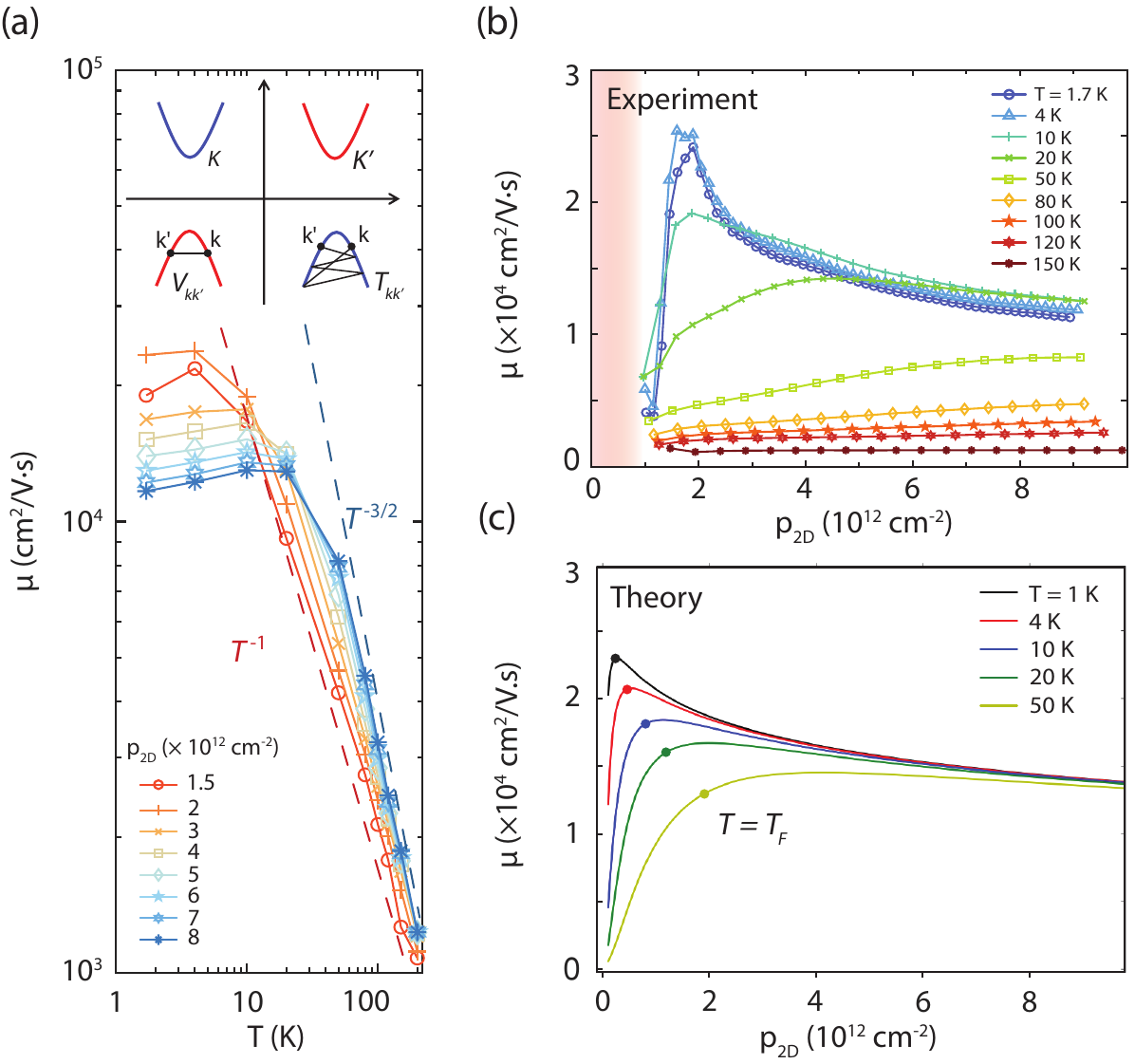}
  \caption{(a) Transport mobility $\mu$ as a function of temperature for fixed hole densities. Dashed lines show $\mu \sim T^{-\gamma}$ for $\gamma = 1, 3/2$ as a guide. (b) Transport mobility $\mu$ as a function of hole density for different temperatures. (c) T-matrix calculations of the mobility as a function of density for various temperatures showing qualitative agreement.
}
\label{fig:3}
\end{figure}

Further evidence for strong short-range scattering potentials are shown in the study of temperature dependent transport mobility. Figure \ref{fig:3}a displays transport mobility plotted as a function of temperature for fixed hole densities. The observed $\mu \sim T^{-\gamma}$ dependence at higher temperatures is a manifestation of optical phonon scattering and the exponent $\gamma$ can be used to characterize the dominating phonon scattering mechanism\cite{Kaasbjerg2017}. At the lowest densities, the mobility scales with $\gamma \sim 1$, indicating acoustic phonon scattering above the Bloch-Gr{\"u}neisen temperature. For higher densities, we observe an increased exponent of $\gamma \sim 3/2$, indicating a transition into optical phonon scattering through the deformation potential couplings and Fr{\"o}hlich interaction\cite{Kaasbjerg2017}. At lower temperatures, however, mobility tends to grow slower as $T$ decreases, due to the diminishing role of electron-phonon scattering over impurity scattering. At this low temperature limit where the impurity scattering becomes appreciable, we find $\mu(T)$ exhibits a complicated behavior, including non-monotonic change with $T$ at the high density limit. The origin of this density dependent anomalous mobility modulation at low temperatures can be related to the scattering rate $\tau^{-1}$ change with density $p_{2D}$ discussed above.  

Figure \ref{fig:3}b shows transport mobility $\mu$ vs hole density at different temperatures, calculated from the measured Hall density $p_{2D}$ and conductivity $\sigma$. $\mu$ reaches the maximum value  $\sim$25,000 cm$^2$/(V$\cdot$s) at 4K and $p \approx p_* \sim 2 \times 10^{12}$ cm$^{-2}$. This high mobility is consistent with optical studies of high-quality WSe$_2$ devices, where in photoluminescence measurements we observe narrow linewidths and emission of complex excitonic states \cite{Li2018,Barbone2018}. As $T$ increases, the density where $\mu$ (and corresponding $\tau$) is peaked, $p_*(T)$, increases rapidly. For higher temperatures, $T>50$~K, we recover the typical mobility vs density dependence, i.e., monotonically decreasing $\mu$ as $p_{2d}$ decreases due to the increasing contribution of unscreened charged defects. We observe a similar mobility vs hole density trend in commercial CVT crystals (see SI), but with an order of magnitude lower mobility of $\sim$3000 cm$^2$/(V$\cdot$s) at 4~K. While the lower mobility is attributed to higher defect density in CVT crystals, the unconventional mobility behavior in both crystals suggests that hole transport in WSe$_2$ monolayers is intrinsically different than in conventional semiconductors that are limited by charged impurities. 

The strong density dependence of the mobility indicates a concomitant break down of the Born approximation for intrinsic defects in 2D semiconducting TMDs. A similar failure of the Born approximation occurs for, e.g., atomic defects in graphene which introduce quasibound defect states near the Dirac point giving rise to resonant scattering and a nontrivial density dependence of the mobility~\cite{Mirlin:Electron,Basko:Resonant,Falko:Adsorbate,Katsnelson:Resonant}. However, in contrast to the situation in graphene, we here find that the density dependence of the mobility in monolayer WSe$_2$ can be traced back to a pronounced \emph{renormalization} of the Born scattering amplitude by 1--2 orders of magnitude due to the strong impurity strength of vacancies~\cite{Kaasbjerg2020}. This is described consistently with the $T$-matrix formalism, where the scattering amplitude acquires an inherent energy dependence. 

For quantitative comparison with experimental data, we perform T-matrix calculations, which can be used to incorporate intrinsic point defects and remote charge impurities (see SI for more detail)\cite{Kaasbjerg2020}. Due to large spin-orbit coupling in the WSe$_2$ valence bands, much larger than the hole filling level, minimal intervalley scattering is expected - we can thus study intrinsic intravalley scattering mechanisms in the transport mobility caused by defects and impurities in the system (Fig. \ref{fig:3}a inset). We find that unconventional increase of the mobility for decreasing density is well captured by short-range impurity scattering from charge neutral point defects. To capture the sharp decrease of the mobility at the lowest densities ($p_{2D}<p_*$), we include remote charge impurities, which become relevant at lower carrier mobilities with less screening. Fig. \ref{fig:3}c shows the calculated mobility as a function of $p_{2D}$ and temperature where we have strong agreement between the theory and experimental results. We estimate an intrinsic impurity density of around $10^{11}$ cm$^{-2}$ for flux grown crystals and $10^{12}$ cm$^{-2}$ for CVT samples, consistent with STM studies of the bulk TMD crystals \cite{Edelberg:2019}. In both cases, the concentration of remote charge (Coulomb) impurities is around $10^{12}$ cm$^{-2}$. These calculations strongly suggest that WSe$_2$ is not limited by charged defects, but rather intrinsic, short-ranged charge-neutral defects in the system. We note that the mobility's dependence on density at low temperatures is consistent in both flux-grown and CVT devices (see SI), implying this is the limiting scattering mechanism for hole transport in WSe$_2$ devices regardless of the defect densities. Density functional theory (DFT) calculations of mid-gap states induced by point defects at either the W or Se sites show a limited density of states (DOS) for Se vacancies \cite{Kaasbjerg2020}. These results suggest that improvements in Se vacancies in WSe$_2$ materials will vastly improve the quality for electric transport applications.

\begin{figure}[!t]
  \centering
  \includegraphics[width=0.99\linewidth]{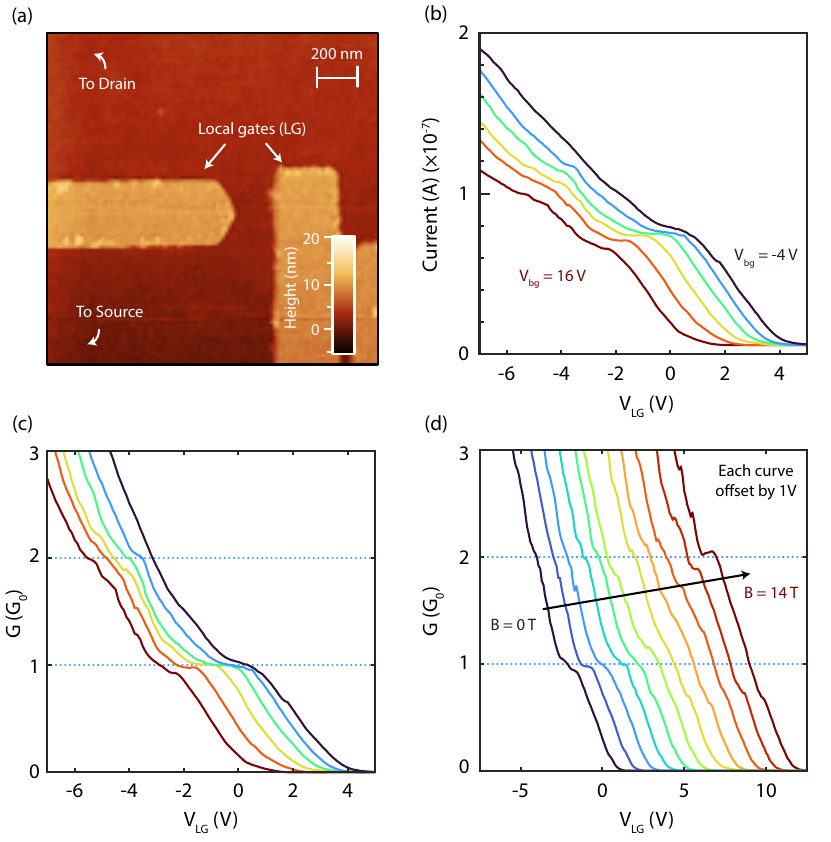}
  \caption{(a) Atomic force microscopy image the local gates ($V_{LG}$) used to form a quantum point contact (QPC). (b) Line cuts of the current as a function of $V_{LG}$ for different V$_{bg}$, showing quantized conductance as the QPC is pinched off. (c) Same linecuts plotted in units of quantum conductance ($G_0$) with subtraction of series resistance ($R_s$), parallel resistance from uncontrolled regions ($R_0$), and $V_{bias}$ reduction due to Schottky barrier effects. (d) Line cuts of conductance as a function of $V_{LG}$ for various $B$ at $V_{bg} = 5V$.
}
\label{fig:4}
\end{figure}

The high mobility we achieve at $p_{2D}=p_*$ correspond to the electron mean free path can reach $\sim$500~nm in monolayer WSe$_2$. This long mean free path enables us to build an electrically controlled quantum point contact (QPC) device using local back gates. In Figure \ref{fig:4}a, we show an atomic force microscopy image of the local gate structure in such a device. A 2D hole gas is generated by electrostatic gating with the contact gates and a global backgate and the potential of the two local gates are shifted together to deplete carriers and create a constriction less than $\sim 200$ nm for quantum confinement. Figure \ref{fig:4}b shows the measured current across the
device as a function of the local-gate voltage ($V_{LG}$) for different back-gate voltages ($V_{bg}$) with a 300 mV alternating current bias. The current exhibits plateau-like features in the current stemming from quantized conductance as the channel width of the QPC approaches the Fermi wavelength. The measured current can be converted to QPC conductance, after subtracting off the series resistances and leakage current in the local gate area (see SI for more detailed procedure). Fig.\ref{fig:4}c shows the QPC conductance corresponding to the data in Fig.\ref{fig:4}b. At least two well-defined conductance plateaus, corresponding to integer steps of $G_0=2e^2/h$, are visible as the QPC constriction becomes wider at lower $V_{LG}$. We find that upon applying perpendicular magnetic fields, these plateaus split (Fig.\ref{fig:4}d). The emergence of two additional conductance steps suggests a lifting of the degenerate spin-locked K valley valence bands at high magnetic fields. 

In conclusion, we report high mobility charge transport in low defect density monolayer WSe$_2$, grown by flux method. We find that while the high temperature mobility is limited by electron-phonon scattering, the low temperature mobility can be controlled by intricate interplay between short and long-range scatters, reaching up to 500 nm electron mean free path at an optimized carrier density . We demonstrate that high quality electronic devices are possible by showing quantized conductance steps from an electrostatically defined quantum point contact, opening a doorway for quantum electronic devices based on monolayer TMDs. 

\begin{acknowledgments}
  P.K. and J.H. acknowledges the support from the Army Research Office’s Multidisciplinary University Initiative (MURI) programme (W911NF-21-2-0147). A.Y.J is supported by Samsung Electronics. K.K. acknowledges support from the European Union's Horizon 2020 research and
  innovation program under the Marie Sklodowska-Curie Grant Agreement
  No.~713683 (COFUNDfellowsDTU). The Center for Nanostructured Graphene (CNG) is
  sponsored by the Danish National Research Foundation, Project DNRF103.K.W. and T.T. acknowledge support from the 310 Elemental Strategy Initiative conducted by the MEXT, Japan, 311 Grant No. JPMXP0112101001, JSPS KAKENHI Grant No. 312 JP20H00354, and the CREST (Grant No. JPMJCR15F3), 313 JST.
\end{acknowledgments}

% $^*$ A.Y.J. and K.P. contributed equally to this work.

%\bibliography{journalabbreviations,references}
% \bibliography{references}

\begin{thebibliography}{38}%
\makeatletter
\providecommand \@ifxundefined [1]{%
 \@ifx{#1\undefined}
}%
\providecommand \@ifnum [1]{%
 \ifnum #1\expandafter \@firstoftwo
 \else \expandafter \@secondoftwo
 \fi
}%
\providecommand \@ifx [1]{%
 \ifx #1\expandafter \@firstoftwo
 \else \expandafter \@secondoftwo
 \fi
}%
\providecommand \natexlab [1]{#1}%
\providecommand \enquote  [1]{``#1''}%
\providecommand \bibnamefont  [1]{#1}%
\providecommand \bibfnamefont [1]{#1}%
\providecommand \citenamefont [1]{#1}%
\providecommand \href@noop [0]{\@secondoftwo}%
\providecommand \href [0]{\begingroup \@sanitize@url \@href}%
\providecommand \@href[1]{\@@startlink{#1}\@@href}%
\providecommand \@@href[1]{\endgroup#1\@@endlink}%
\providecommand \@sanitize@url [0]{\catcode `\\12\catcode `\$12\catcode
  `\&12\catcode `\#12\catcode `\^12\catcode `\_12\catcode `\%12\relax}%
\providecommand \@@startlink[1]{}%
\providecommand \@@endlink[0]{}%
\providecommand \url  [0]{\begingroup\@sanitize@url \@url }%
\providecommand \@url [1]{\endgroup\@href {#1}{\urlprefix }}%
\providecommand \urlprefix  [0]{URL }%
\providecommand \Eprint [0]{\href }%
\providecommand \doibase [0]{http://dx.doi.org/}%
\providecommand \selectlanguage [0]{\@gobble}%
\providecommand \bibinfo  [0]{\@secondoftwo}%
\providecommand \bibfield  [0]{\@secondoftwo}%
\providecommand \translation [1]{[#1]}%
\providecommand \BibitemOpen [0]{}%
\providecommand \bibitemStop [0]{}%
\providecommand \bibitemNoStop [0]{.\EOS\space}%
\providecommand \EOS [0]{\spacefactor3000\relax}%
\providecommand \BibitemShut  [1]{\csname bibitem#1\endcsname}%
\let\auto@bib@innerbib\@empty
%</preamble>
\bibitem [{\citenamefont {Mak}\ \emph {et~al.}(2010)\citenamefont {Mak},
  \citenamefont {Lee}, \citenamefont {Hone}, \citenamefont {Shan},\ and\
  \citenamefont {Heinz}}]{Heinz:ThinMoS2}%
  \BibitemOpen
  \bibfield  {author} {\bibinfo {author} {\bibfnamefont {K.~F.}\ \bibnamefont
  {Mak}}, \bibinfo {author} {\bibfnamefont {C.}~\bibnamefont {Lee}}, \bibinfo
  {author} {\bibfnamefont {J.}~\bibnamefont {Hone}}, \bibinfo {author}
  {\bibfnamefont {J.}~\bibnamefont {Shan}}, \ and\ \bibinfo {author}
  {\bibfnamefont {T.~F.}\ \bibnamefont {Heinz}},\ }\href@noop {} {\bibfield
  {journal} {\bibinfo  {journal} {Phys. Rev. Lett.}\ }\textbf {\bibinfo
  {volume} {105}},\ \bibinfo {pages} {136805} (\bibinfo {year}
  {2010})}\BibitemShut {NoStop}%
\bibitem [{\citenamefont {Radisavljevic}\ \emph {et~al.}(2011)\citenamefont
  {Radisavljevic}, \citenamefont {Radenovic}, \citenamefont {Brivio},
  \citenamefont {Giacometti},\ and\ \citenamefont {Kis}}]{Kis:MoS2Transistor}%
  \BibitemOpen
  \bibfield  {author} {\bibinfo {author} {\bibfnamefont {B.}~\bibnamefont
  {Radisavljevic}}, \bibinfo {author} {\bibfnamefont {A.}~\bibnamefont
  {Radenovic}}, \bibinfo {author} {\bibfnamefont {J.}~\bibnamefont {Brivio}},
  \bibinfo {author} {\bibfnamefont {V.}~\bibnamefont {Giacometti}}, \ and\
  \bibinfo {author} {\bibfnamefont {A.}~\bibnamefont {Kis}},\ }\href@noop {}
  {\bibfield  {journal} {\bibinfo  {journal} {Nature Nano.}\ }\textbf {\bibinfo
  {volume} {6}},\ \bibinfo {pages} {147} (\bibinfo {year} {2011})}\BibitemShut
  {NoStop}%
\bibitem [{\citenamefont {Korn}\ \emph {et~al.}(2011)\citenamefont {Korn},
  \citenamefont {Heydrich}, \citenamefont {Hirmer}, \citenamefont
  {Schmutzler},\ and\ \citenamefont {Sch{\"u}ller}}]{Schuller:Photocarrier}%
  \BibitemOpen
  \bibfield  {author} {\bibinfo {author} {\bibfnamefont {T.}~\bibnamefont
  {Korn}}, \bibinfo {author} {\bibfnamefont {S.}~\bibnamefont {Heydrich}},
  \bibinfo {author} {\bibfnamefont {M.}~\bibnamefont {Hirmer}}, \bibinfo
  {author} {\bibfnamefont {J.}~\bibnamefont {Schmutzler}}, \ and\ \bibinfo
  {author} {\bibfnamefont {C.}~\bibnamefont {Sch{\"u}ller}},\ }\href@noop {}
  {\bibfield  {journal} {\bibinfo  {journal} {Appl. Phys. Lett.}\ }\textbf
  {\bibinfo {volume} {99}},\ \bibinfo {pages} {102109} (\bibinfo {year}
  {2011})}\BibitemShut {NoStop}%
\bibitem [{\citenamefont {Avouris}\ \emph {et~al.}(2017)\citenamefont
  {Avouris}, \citenamefont {Heinz},\ and\ \citenamefont {Low}}]{avouris20172d}%
  \BibitemOpen
  \bibfield  {author} {\bibinfo {author} {\bibfnamefont {P.}~\bibnamefont
  {Avouris}}, \bibinfo {author} {\bibfnamefont {T.~F.}\ \bibnamefont {Heinz}},
  \ and\ \bibinfo {author} {\bibfnamefont {T.}~\bibnamefont {Low}},\
  }\href@noop {} {\emph {\bibinfo {title} {2D Materials}}}\ (\bibinfo
  {publisher} {Cambridge University Press},\ \bibinfo {year}
  {2017})\BibitemShut {NoStop}%
\bibitem [{\citenamefont {Xiao}\ \emph {et~al.}(2012)\citenamefont {Xiao},
  \citenamefont {Liu}, \citenamefont {Feng}, \citenamefont {Xu},\ and\
  \citenamefont {Yao}}]{Yao:SpinValley}%
  \BibitemOpen
  \bibfield  {author} {\bibinfo {author} {\bibfnamefont {D.}~\bibnamefont
  {Xiao}}, \bibinfo {author} {\bibfnamefont {G.-B.}\ \bibnamefont {Liu}},
  \bibinfo {author} {\bibfnamefont {W.}~\bibnamefont {Feng}}, \bibinfo {author}
  {\bibfnamefont {X.}~\bibnamefont {Xu}}, \ and\ \bibinfo {author}
  {\bibfnamefont {W.}~\bibnamefont {Yao}},\ }\href@noop {} {\bibfield
  {journal} {\bibinfo  {journal} {Phys. Rev. Lett.}\ }\textbf {\bibinfo
  {volume} {108}},\ \bibinfo {pages} {196802} (\bibinfo {year}
  {2012})}\BibitemShut {NoStop}%
\bibitem [{\citenamefont {Yao}\ \emph {et~al.}(2014)\citenamefont {Yao},
  \citenamefont {Xiao},\ and\ \citenamefont {Heinz}}]{Heinz:Spin}%
  \BibitemOpen
  \bibfield  {author} {\bibinfo {author} {\bibfnamefont {X.~X.~W.}\
  \bibnamefont {Yao}}, \bibinfo {author} {\bibfnamefont {D.}~\bibnamefont
  {Xiao}}, \ and\ \bibinfo {author} {\bibfnamefont {T.~F.}\ \bibnamefont
  {Heinz}},\ }\href@noop {} {\bibfield  {journal} {\bibinfo  {journal} {Nature
  Phys.}\ }\textbf {\bibinfo {volume} {10}},\ \bibinfo {pages} {343} (\bibinfo
  {year} {2014})}\BibitemShut {NoStop}%
\bibitem [{\citenamefont {Dean}\ \emph {et~al.}(2010)\citenamefont {Dean},
  \citenamefont {Young}, \citenamefont {Meric}, \citenamefont {Lee},
  \citenamefont {Wang}, \citenamefont {Sorgenfrei}, \citenamefont {Watanabe},
  \citenamefont {Taniguchi}, \citenamefont {Kim}, \citenamefont {Shepard},\
  and\ \citenamefont {Hone}}]{Dean2010}%
  \BibitemOpen
  \bibfield  {author} {\bibinfo {author} {\bibfnamefont {C.~R.}\ \bibnamefont
  {Dean}}, \bibinfo {author} {\bibfnamefont {A.~F.}\ \bibnamefont {Young}},
  \bibinfo {author} {\bibfnamefont {I.}~\bibnamefont {Meric}}, \bibinfo
  {author} {\bibfnamefont {C.}~\bibnamefont {Lee}}, \bibinfo {author}
  {\bibfnamefont {L.}~\bibnamefont {Wang}}, \bibinfo {author} {\bibfnamefont
  {S.}~\bibnamefont {Sorgenfrei}}, \bibinfo {author} {\bibfnamefont
  {K.}~\bibnamefont {Watanabe}}, \bibinfo {author} {\bibfnamefont
  {T.}~\bibnamefont {Taniguchi}}, \bibinfo {author} {\bibfnamefont
  {P.}~\bibnamefont {Kim}}, \bibinfo {author} {\bibfnamefont {K.~L.}\
  \bibnamefont {Shepard}}, \ and\ \bibinfo {author} {\bibfnamefont
  {J.}~\bibnamefont {Hone}},\ }\href {\doibase 10.1038/nnano.2010.172}
  {\bibfield  {journal} {\bibinfo  {journal} {Nature Nanotechnology}\ }\textbf
  {\bibinfo {volume} {5}},\ \bibinfo {pages} {722} (\bibinfo {year} {2010})},\
  \Eprint {http://arxiv.org/abs/1005.4917} {1005.4917} \BibitemShut {NoStop}%
\bibitem [{\citenamefont {Chung}\ \emph {et~al.}(2021)\citenamefont {Chung},
  \citenamefont {Rosales}, \citenamefont {Baldwin}, \citenamefont {Madathil},
  \citenamefont {West}, \citenamefont {Shayegan},\ and\ \citenamefont
  {Pfeiffer}}]{Chung2021}%
  \BibitemOpen
  \bibfield  {author} {\bibinfo {author} {\bibfnamefont {Y.~J.}\ \bibnamefont
  {Chung}}, \bibinfo {author} {\bibfnamefont {K.~A.~V.}\ \bibnamefont
  {Rosales}}, \bibinfo {author} {\bibfnamefont {K.~W.}\ \bibnamefont
  {Baldwin}}, \bibinfo {author} {\bibfnamefont {P.~T.}\ \bibnamefont
  {Madathil}}, \bibinfo {author} {\bibfnamefont {K.~W.}\ \bibnamefont {West}},
  \bibinfo {author} {\bibfnamefont {M.}~\bibnamefont {Shayegan}}, \ and\
  \bibinfo {author} {\bibfnamefont {L.~N.}\ \bibnamefont {Pfeiffer}},\
  }\href@noop {} {\bibfield  {journal} {\bibinfo  {journal} {Nature Materials}\
  }\textbf {\bibinfo {volume} {20}},\ \bibinfo {pages} {632–637} (\bibinfo
  {year} {2021})}\BibitemShut {NoStop}%
\bibitem [{\citenamefont {Radisavljevic}\ and\ \citenamefont
  {Kis}(2013)}]{Kis:Engineering}%
  \BibitemOpen
  \bibfield  {author} {\bibinfo {author} {\bibfnamefont {B.}~\bibnamefont
  {Radisavljevic}}\ and\ \bibinfo {author} {\bibfnamefont {A.}~\bibnamefont
  {Kis}},\ }\href@noop {} {\bibfield  {journal} {\bibinfo  {journal} {Nature
  Mat.}\ }\textbf {\bibinfo {volume} {12}},\ \bibinfo {pages} {815} (\bibinfo
  {year} {2013})}\BibitemShut {NoStop}%
\bibitem [{\citenamefont {Baugher}\ \emph {et~al.}(2013)\citenamefont
  {Baugher}, \citenamefont {Churchill}, \citenamefont {Yang},\ and\
  \citenamefont {Jarillo-Herrero}}]{Herrero:Intrinsic}%
  \BibitemOpen
  \bibfield  {author} {\bibinfo {author} {\bibfnamefont {B.~W.~H.}\
  \bibnamefont {Baugher}}, \bibinfo {author} {\bibfnamefont {H.~O.~H.}\
  \bibnamefont {Churchill}}, \bibinfo {author} {\bibfnamefont {Y.}~\bibnamefont
  {Yang}}, \ and\ \bibinfo {author} {\bibfnamefont {P.}~\bibnamefont
  {Jarillo-Herrero}},\ }\href@noop {} {\bibfield  {journal} {\bibinfo
  {journal} {Nano. Lett.}\ }\textbf {\bibinfo {volume} {13}},\ \bibinfo {pages}
  {4212} (\bibinfo {year} {2013})}\BibitemShut {NoStop}%
\bibitem [{\citenamefont {Zhu}\ \emph {et~al.}(2014)\citenamefont {Zhu},
  \citenamefont {Low}, \citenamefont {Lee}, \citenamefont {Wang}, \citenamefont
  {Farmer}, \citenamefont {Kong}, \citenamefont {Xia},\ and\ \citenamefont
  {Avouris}}]{Avouris:Electronic}%
  \BibitemOpen
  \bibfield  {author} {\bibinfo {author} {\bibfnamefont {W.}~\bibnamefont
  {Zhu}}, \bibinfo {author} {\bibfnamefont {T.}~\bibnamefont {Low}}, \bibinfo
  {author} {\bibfnamefont {Y.-H.}\ \bibnamefont {Lee}}, \bibinfo {author}
  {\bibfnamefont {H.}~\bibnamefont {Wang}}, \bibinfo {author} {\bibfnamefont
  {D.~B.}\ \bibnamefont {Farmer}}, \bibinfo {author} {\bibfnamefont
  {J.}~\bibnamefont {Kong}}, \bibinfo {author} {\bibfnamefont {F.}~\bibnamefont
  {Xia}}, \ and\ \bibinfo {author} {\bibfnamefont {P.}~\bibnamefont
  {Avouris}},\ }\href@noop {} {\bibfield  {journal} {\bibinfo  {journal}
  {Nature Commun.}\ }\textbf {\bibinfo {volume} {5}},\ \bibinfo {pages} {3087}
  (\bibinfo {year} {2014})}\BibitemShut {NoStop}%
\bibitem [{\citenamefont {Yu}\ \emph {et~al.}(2014)\citenamefont {Yu},
  \citenamefont {Pan}, \citenamefont {Shen}, \citenamefont {Wang},
  \citenamefont {Ong}, \citenamefont {Xu}, \citenamefont {Xin}, \citenamefont
  {Pan}, \citenamefont {Wang}, \citenamefont {Sun}, \citenamefont {Wang},
  \citenamefont {Zhang}, \citenamefont {Zhang}, \citenamefont {Shi},\ and\
  \citenamefont {Wang}}]{Wang:Towards}%
  \BibitemOpen
  \bibfield  {author} {\bibinfo {author} {\bibfnamefont {Z.}~\bibnamefont
  {Yu}}, \bibinfo {author} {\bibfnamefont {Y.}~\bibnamefont {Pan}}, \bibinfo
  {author} {\bibfnamefont {Y.}~\bibnamefont {Shen}}, \bibinfo {author}
  {\bibfnamefont {Z.}~\bibnamefont {Wang}}, \bibinfo {author} {\bibfnamefont
  {Z.-Y.}\ \bibnamefont {Ong}}, \bibinfo {author} {\bibfnamefont
  {T.}~\bibnamefont {Xu}}, \bibinfo {author} {\bibfnamefont {R.}~\bibnamefont
  {Xin}}, \bibinfo {author} {\bibfnamefont {L.}~\bibnamefont {Pan}}, \bibinfo
  {author} {\bibfnamefont {B.}~\bibnamefont {Wang}}, \bibinfo {author}
  {\bibfnamefont {L.}~\bibnamefont {Sun}}, \bibinfo {author} {\bibfnamefont
  {J.}~\bibnamefont {Wang}}, \bibinfo {author} {\bibfnamefont {G.}~\bibnamefont
  {Zhang}}, \bibinfo {author} {\bibfnamefont {Y.~W.}\ \bibnamefont {Zhang}},
  \bibinfo {author} {\bibfnamefont {Y.}~\bibnamefont {Shi}}, \ and\ \bibinfo
  {author} {\bibfnamefont {X.}~\bibnamefont {Wang}},\ }\href@noop {} {\bibfield
   {journal} {\bibinfo  {journal} {Nature Commun.}\ }\textbf {\bibinfo {volume}
  {5}},\ \bibinfo {pages} {5290} (\bibinfo {year} {2014})}\BibitemShut
  {NoStop}%
\bibitem [{\citenamefont {Schmidt}\ \emph {et~al.}(2014)\citenamefont
  {Schmidt}, \citenamefont {Wang}, \citenamefont {Chu}, \citenamefont {Toh},
  \citenamefont {Kumar}, \citenamefont {Zhao}, \citenamefont {Neto},
  \citenamefont {Martin}, \citenamefont {Adam}, \citenamefont {{\"O}zyilmaz},\
  and\ \citenamefont {Eda}}]{Eda:Transport}%
  \BibitemOpen
  \bibfield  {author} {\bibinfo {author} {\bibfnamefont {H.}~\bibnamefont
  {Schmidt}}, \bibinfo {author} {\bibfnamefont {S.}~\bibnamefont {Wang}},
  \bibinfo {author} {\bibfnamefont {L.}~\bibnamefont {Chu}}, \bibinfo {author}
  {\bibfnamefont {M.}~\bibnamefont {Toh}}, \bibinfo {author} {\bibfnamefont
  {R.}~\bibnamefont {Kumar}}, \bibinfo {author} {\bibfnamefont
  {W.}~\bibnamefont {Zhao}}, \bibinfo {author} {\bibfnamefont {A.~H.~C.}\
  \bibnamefont {Neto}}, \bibinfo {author} {\bibfnamefont {J.}~\bibnamefont
  {Martin}}, \bibinfo {author} {\bibfnamefont {S.}~\bibnamefont {Adam}},
  \bibinfo {author} {\bibfnamefont {B.}~\bibnamefont {{\"O}zyilmaz}}, \ and\
  \bibinfo {author} {\bibfnamefont {G.}~\bibnamefont {Eda}},\ }\href@noop {}
  {\bibfield  {journal} {\bibinfo  {journal} {Nano. Lett.}\ }\textbf {\bibinfo
  {volume} {14}},\ \bibinfo {pages} {1909} (\bibinfo {year}
  {2014})}\BibitemShut {NoStop}%
\bibitem [{\citenamefont {Chu}\ \emph {et~al.}(2014)\citenamefont {Chu},
  \citenamefont {Schmidt}, \citenamefont {Pu}, \citenamefont {Wang},
  \citenamefont {{\"O}zyilmaz}, \citenamefont {Takenobu},\ and\ \citenamefont
  {Eda}}]{Eda:Charge}%
  \BibitemOpen
  \bibfield  {author} {\bibinfo {author} {\bibfnamefont {L.}~\bibnamefont
  {Chu}}, \bibinfo {author} {\bibfnamefont {H.}~\bibnamefont {Schmidt}},
  \bibinfo {author} {\bibfnamefont {J.}~\bibnamefont {Pu}}, \bibinfo {author}
  {\bibfnamefont {S.}~\bibnamefont {Wang}}, \bibinfo {author} {\bibfnamefont
  {B.}~\bibnamefont {{\"O}zyilmaz}}, \bibinfo {author} {\bibfnamefont
  {T.}~\bibnamefont {Takenobu}}, \ and\ \bibinfo {author} {\bibfnamefont
  {G.}~\bibnamefont {Eda}},\ }\href@noop {} {\bibfield  {journal} {\bibinfo
  {journal} {Scientific Reports}\ }\textbf {\bibinfo {volume} {4}},\ \bibinfo
  {pages} {7293} (\bibinfo {year} {2014})}\BibitemShut {NoStop}%
\bibitem [{\citenamefont {Cui}\ \emph {et~al.}(2015)\citenamefont {Cui},
  \citenamefont {Lee}, \citenamefont {Kim}, \citenamefont {Arefe},
  \citenamefont {Huang}, \citenamefont {Lee}, \citenamefont {Chenet},
  \citenamefont {Zhang}, \citenamefont {Wang}, \citenamefont {Ye},
  \citenamefont {Pizzocchero}, \citenamefont {Jessen}, \citenamefont
  {Watanabe}, \citenamefont {Taniguchi}, \citenamefont {Muller}, \citenamefont
  {Low}, \citenamefont {Kim},\ and\ \citenamefont {Hone}}]{Hone:Multi}%
  \BibitemOpen
  \bibfield  {author} {\bibinfo {author} {\bibfnamefont {X.}~\bibnamefont
  {Cui}}, \bibinfo {author} {\bibfnamefont {G.-H.}\ \bibnamefont {Lee}},
  \bibinfo {author} {\bibfnamefont {Y.~D.}\ \bibnamefont {Kim}}, \bibinfo
  {author} {\bibfnamefont {G.}~\bibnamefont {Arefe}}, \bibinfo {author}
  {\bibfnamefont {P.~Y.}\ \bibnamefont {Huang}}, \bibinfo {author}
  {\bibfnamefont {C.-H.}\ \bibnamefont {Lee}}, \bibinfo {author} {\bibfnamefont
  {D.~A.}\ \bibnamefont {Chenet}}, \bibinfo {author} {\bibfnamefont
  {X.}~\bibnamefont {Zhang}}, \bibinfo {author} {\bibfnamefont
  {L.}~\bibnamefont {Wang}}, \bibinfo {author} {\bibfnamefont {F.}~\bibnamefont
  {Ye}}, \bibinfo {author} {\bibfnamefont {F.}~\bibnamefont {Pizzocchero}},
  \bibinfo {author} {\bibfnamefont {B.~S.}\ \bibnamefont {Jessen}}, \bibinfo
  {author} {\bibfnamefont {K.}~\bibnamefont {Watanabe}}, \bibinfo {author}
  {\bibfnamefont {T.}~\bibnamefont {Taniguchi}}, \bibinfo {author}
  {\bibfnamefont {D.~A.}\ \bibnamefont {Muller}}, \bibinfo {author}
  {\bibfnamefont {T.}~\bibnamefont {Low}}, \bibinfo {author} {\bibfnamefont
  {P.}~\bibnamefont {Kim}}, \ and\ \bibinfo {author} {\bibfnamefont
  {J.}~\bibnamefont {Hone}},\ }\href@noop {} {\bibfield  {journal} {\bibinfo
  {journal} {Nature Nano.}\ }\textbf {\bibinfo {volume} {10}},\ \bibinfo
  {pages} {534} (\bibinfo {year} {2015})}\BibitemShut {NoStop}%
\bibitem [{\citenamefont {Schmidt}\ \emph {et~al.}(2016)\citenamefont
  {Schmidt}, \citenamefont {Yudhistira}, \citenamefont {Chu}, \citenamefont
  {Neto}, \citenamefont {{\"O}zyilmaz}, \citenamefont {Adam},\ and\
  \citenamefont {Eda}}]{Eda:Quantum}%
  \BibitemOpen
  \bibfield  {author} {\bibinfo {author} {\bibfnamefont {H.}~\bibnamefont
  {Schmidt}}, \bibinfo {author} {\bibfnamefont {I.}~\bibnamefont {Yudhistira}},
  \bibinfo {author} {\bibfnamefont {L.}~\bibnamefont {Chu}}, \bibinfo {author}
  {\bibfnamefont {A.~H.~C.}\ \bibnamefont {Neto}}, \bibinfo {author}
  {\bibfnamefont {B.}~\bibnamefont {{\"O}zyilmaz}}, \bibinfo {author}
  {\bibfnamefont {S.}~\bibnamefont {Adam}}, \ and\ \bibinfo {author}
  {\bibfnamefont {G.}~\bibnamefont {Eda}},\ }\href@noop {} {\bibfield
  {journal} {\bibinfo  {journal} {Phys. Rev. Lett.}\ }\textbf {\bibinfo
  {volume} {116}},\ \bibinfo {pages} {046803} (\bibinfo {year}
  {2016})}\BibitemShut {NoStop}%
\bibitem [{\citenamefont {Cui}\ \emph {et~al.}(2017)\citenamefont {Cui},
  \citenamefont {Shih}, \citenamefont {Jauregui}, \citenamefont {Chae},
  \citenamefont {Kim}, \citenamefont {Li}, \citenamefont {Seo}, \citenamefont
  {Pistunova}, \citenamefont {Yin}, \citenamefont {Park}, \citenamefont {Choi},
  \citenamefont {Lee}, \citenamefont {Watanabe}, \citenamefont {Taniguchi},
  \citenamefont {Kim}, \citenamefont {Dean},\ and\ \citenamefont
  {Hone}}]{Hone:Low}%
  \BibitemOpen
  \bibfield  {author} {\bibinfo {author} {\bibfnamefont {X.}~\bibnamefont
  {Cui}}, \bibinfo {author} {\bibfnamefont {E.-M.}\ \bibnamefont {Shih}},
  \bibinfo {author} {\bibfnamefont {L.~A.}\ \bibnamefont {Jauregui}}, \bibinfo
  {author} {\bibfnamefont {S.~H.}\ \bibnamefont {Chae}}, \bibinfo {author}
  {\bibfnamefont {Y.~D.}\ \bibnamefont {Kim}}, \bibinfo {author} {\bibfnamefont
  {B.}~\bibnamefont {Li}}, \bibinfo {author} {\bibfnamefont {D.}~\bibnamefont
  {Seo}}, \bibinfo {author} {\bibfnamefont {K.}~\bibnamefont {Pistunova}},
  \bibinfo {author} {\bibfnamefont {J.}~\bibnamefont {Yin}}, \bibinfo {author}
  {\bibfnamefont {J.-H.}\ \bibnamefont {Park}}, \bibinfo {author}
  {\bibfnamefont {H.-J.}\ \bibnamefont {Choi}}, \bibinfo {author}
  {\bibfnamefont {Y.~H.}\ \bibnamefont {Lee}}, \bibinfo {author} {\bibfnamefont
  {K.}~\bibnamefont {Watanabe}}, \bibinfo {author} {\bibfnamefont
  {T.}~\bibnamefont {Taniguchi}}, \bibinfo {author} {\bibfnamefont
  {P.}~\bibnamefont {Kim}}, \bibinfo {author} {\bibfnamefont {C.~R.}\
  \bibnamefont {Dean}}, \ and\ \bibinfo {author} {\bibfnamefont {J.~C.}\
  \bibnamefont {Hone}},\ }\href@noop {} {\bibfield  {journal} {\bibinfo
  {journal} {Nano. Lett.}\ }\textbf {\bibinfo {volume} {17}},\ \bibinfo {pages}
  {4781} (\bibinfo {year} {2017})}\BibitemShut {NoStop}%
\bibitem [{\citenamefont {Fallahazad}\ \emph {et~al.}(2016)\citenamefont
  {Fallahazad}, \citenamefont {Movva}, \citenamefont {Kim}, \citenamefont
  {Larentis}, \citenamefont {Taniguchi}, \citenamefont {Watanabe},
  \citenamefont {Banerjee},\ and\ \citenamefont {Tutuc}}]{Tutuc:Contacts}%
  \BibitemOpen
  \bibfield  {author} {\bibinfo {author} {\bibfnamefont {B.}~\bibnamefont
  {Fallahazad}}, \bibinfo {author} {\bibfnamefont {H.~C.}\ \bibnamefont
  {Movva}}, \bibinfo {author} {\bibfnamefont {K.}~\bibnamefont {Kim}}, \bibinfo
  {author} {\bibfnamefont {S.}~\bibnamefont {Larentis}}, \bibinfo {author}
  {\bibfnamefont {T.}~\bibnamefont {Taniguchi}}, \bibinfo {author}
  {\bibfnamefont {K.}~\bibnamefont {Watanabe}}, \bibinfo {author}
  {\bibfnamefont {S.~K.}\ \bibnamefont {Banerjee}}, \ and\ \bibinfo {author}
  {\bibfnamefont {E.}~\bibnamefont {Tutuc}},\ }\href {\doibase
  10.1103/PhysRevLett.116.086601} {\bibfield  {journal} {\bibinfo  {journal}
  {Physical Review Letters}\ }\textbf {\bibinfo {volume} {116}},\ \bibinfo
  {pages} {1} (\bibinfo {year} {2016})}\BibitemShut {NoStop}%
\bibitem [{\citenamefont {Pisoni}\ \emph
  {et~al.}(2018{\natexlab{a}})\citenamefont {Pisoni}, \citenamefont {Lei},
  \citenamefont {Back}, \citenamefont {Eich}, \citenamefont {Overweg},
  \citenamefont {Lee}, \citenamefont {Watanabe}, \citenamefont {Taniguchi},
  \citenamefont {Ihn},\ and\ \citenamefont {Ensslin}}]{Ensslin:Gate}%
  \BibitemOpen
  \bibfield  {author} {\bibinfo {author} {\bibfnamefont {R.}~\bibnamefont
  {Pisoni}}, \bibinfo {author} {\bibfnamefont {Z.}~\bibnamefont {Lei}},
  \bibinfo {author} {\bibfnamefont {P.}~\bibnamefont {Back}}, \bibinfo {author}
  {\bibfnamefont {M.}~\bibnamefont {Eich}}, \bibinfo {author} {\bibfnamefont
  {H.}~\bibnamefont {Overweg}}, \bibinfo {author} {\bibfnamefont
  {Y.}~\bibnamefont {Lee}}, \bibinfo {author} {\bibfnamefont {K.}~\bibnamefont
  {Watanabe}}, \bibinfo {author} {\bibfnamefont {T.}~\bibnamefont {Taniguchi}},
  \bibinfo {author} {\bibfnamefont {T.}~\bibnamefont {Ihn}}, \ and\ \bibinfo
  {author} {\bibfnamefont {K.}~\bibnamefont {Ensslin}},\ }\href@noop {}
  {\bibfield  {journal} {\bibinfo  {journal} {Appl. Phys. Lett.}\ }\textbf
  {\bibinfo {volume} {112}},\ \bibinfo {pages} {123101} (\bibinfo {year}
  {2018}{\natexlab{a}})}\BibitemShut {NoStop}%
\bibitem [{\citenamefont {Gustafsson}\ \emph {et~al.}(2018)\citenamefont
  {Gustafsson}, \citenamefont {Yankowitz}, \citenamefont {Forsythe},
  \citenamefont {Rhodes}, \citenamefont {Watanabe}, \citenamefont {Taniguchi},
  \citenamefont {Hone}, \citenamefont {Zhu},\ and\ \citenamefont
  {Dean}}]{Gustafsson2018}%
  \BibitemOpen
  \bibfield  {author} {\bibinfo {author} {\bibfnamefont {M.~V.}\ \bibnamefont
  {Gustafsson}}, \bibinfo {author} {\bibfnamefont {M.}~\bibnamefont
  {Yankowitz}}, \bibinfo {author} {\bibfnamefont {C.}~\bibnamefont {Forsythe}},
  \bibinfo {author} {\bibfnamefont {D.}~\bibnamefont {Rhodes}}, \bibinfo
  {author} {\bibfnamefont {K.}~\bibnamefont {Watanabe}}, \bibinfo {author}
  {\bibfnamefont {T.}~\bibnamefont {Taniguchi}}, \bibinfo {author}
  {\bibfnamefont {J.}~\bibnamefont {Hone}}, \bibinfo {author} {\bibfnamefont
  {X.}~\bibnamefont {Zhu}}, \ and\ \bibinfo {author} {\bibfnamefont {C.~R.}\
  \bibnamefont {Dean}},\ }\href@noop {} {\bibfield  {journal} {\bibinfo
  {journal} {Nature Mat.}\ }\textbf {\bibinfo {volume} {17}},\ \bibinfo {pages}
  {411} (\bibinfo {year} {2018})}\BibitemShut {NoStop}%
\bibitem [{\citenamefont {Larentis}\ \emph {et~al.}(2018)\citenamefont
  {Larentis}, \citenamefont {Movva}, \citenamefont {Fallahazad}, \citenamefont
  {Kim}, \citenamefont {Behroozi}, \citenamefont {Taniguchi}, \citenamefont
  {Watanabe}, \citenamefont {Banerjee},\ and\ \citenamefont
  {Tutuc}}]{Tutuc:Large}%
  \BibitemOpen
  \bibfield  {author} {\bibinfo {author} {\bibfnamefont {S.}~\bibnamefont
  {Larentis}}, \bibinfo {author} {\bibfnamefont {H.~C.~P.}\ \bibnamefont
  {Movva}}, \bibinfo {author} {\bibfnamefont {B.}~\bibnamefont {Fallahazad}},
  \bibinfo {author} {\bibfnamefont {K.}~\bibnamefont {Kim}}, \bibinfo {author}
  {\bibfnamefont {A.}~\bibnamefont {Behroozi}}, \bibinfo {author}
  {\bibfnamefont {T.}~\bibnamefont {Taniguchi}}, \bibinfo {author}
  {\bibfnamefont {K.}~\bibnamefont {Watanabe}}, \bibinfo {author}
  {\bibfnamefont {S.~K.}\ \bibnamefont {Banerjee}}, \ and\ \bibinfo {author}
  {\bibfnamefont {E.}~\bibnamefont {Tutuc}},\ }\href@noop {} {\bibfield
  {journal} {\bibinfo  {journal} {Phys. Rev. B}\ }\textbf {\bibinfo {volume}
  {97}},\ \bibinfo {pages} {201407(R)} (\bibinfo {year} {2018})}\BibitemShut
  {NoStop}%
\bibitem [{\citenamefont {Pisoni}\ \emph
  {et~al.}(2018{\natexlab{b}})\citenamefont {Pisoni}, \citenamefont
  {Korm\'anyos}, \citenamefont {Brooks}, \citenamefont {Lei}, \citenamefont
  {Back}, \citenamefont {Eich}, \citenamefont {Overweg}, \citenamefont {Lee},
  \citenamefont {Rickhaus}, \citenamefont {Watanabe}, \citenamefont
  {Taniguchi}, \citenamefont {Imamoglu}, \citenamefont {Burkard}, \citenamefont
  {Ihn},\ and\ \citenamefont {Ensslin}}]{Ensslin:Interactions}%
  \BibitemOpen
  \bibfield  {author} {\bibinfo {author} {\bibfnamefont {R.}~\bibnamefont
  {Pisoni}}, \bibinfo {author} {\bibfnamefont {A.}~\bibnamefont {Korm\'anyos}},
  \bibinfo {author} {\bibfnamefont {M.}~\bibnamefont {Brooks}}, \bibinfo
  {author} {\bibfnamefont {Z.}~\bibnamefont {Lei}}, \bibinfo {author}
  {\bibfnamefont {P.}~\bibnamefont {Back}}, \bibinfo {author} {\bibfnamefont
  {M.}~\bibnamefont {Eich}}, \bibinfo {author} {\bibfnamefont {H.}~\bibnamefont
  {Overweg}}, \bibinfo {author} {\bibfnamefont {Y.}~\bibnamefont {Lee}},
  \bibinfo {author} {\bibfnamefont {P.}~\bibnamefont {Rickhaus}}, \bibinfo
  {author} {\bibfnamefont {K.}~\bibnamefont {Watanabe}}, \bibinfo {author}
  {\bibfnamefont {T.}~\bibnamefont {Taniguchi}}, \bibinfo {author}
  {\bibfnamefont {A.}~\bibnamefont {Imamoglu}}, \bibinfo {author}
  {\bibfnamefont {G.}~\bibnamefont {Burkard}}, \bibinfo {author} {\bibfnamefont
  {T.}~\bibnamefont {Ihn}}, \ and\ \bibinfo {author} {\bibfnamefont
  {K.}~\bibnamefont {Ensslin}},\ }\href {\doibase
  10.1103/PhysRevLett.121.247701} {\bibfield  {journal} {\bibinfo  {journal}
  {Phys. Rev. Lett.}\ }\textbf {\bibinfo {volume} {121}},\ \bibinfo {pages}
  {247701} (\bibinfo {year} {2018}{\natexlab{b}})}\BibitemShut {NoStop}%
\bibitem [{\citenamefont {Edelberg}\ \emph {et~al.}(2019)\citenamefont
  {Edelberg}, \citenamefont {Rhodes}, \citenamefont {Kerelsky}, \citenamefont
  {Kim}, \citenamefont {Wang}, \citenamefont {Zangiabadi}, \citenamefont {Kim},
  \citenamefont {Abhinandan}, \citenamefont {Ardelean}, \citenamefont {Scully},
  \citenamefont {Scullion}, \citenamefont {Embon}, \citenamefont {Zu},
  \citenamefont {Santos}, \citenamefont {Balicas}, \citenamefont {Marianetti},
  \citenamefont {Barmak}, \citenamefont {Zhu}, \citenamefont {Hone},\ and\
  \citenamefont {Pasupathy}}]{Edelberg:2019}%
  \BibitemOpen
  \bibfield  {author} {\bibinfo {author} {\bibfnamefont {D.}~\bibnamefont
  {Edelberg}}, \bibinfo {author} {\bibfnamefont {D.}~\bibnamefont {Rhodes}},
  \bibinfo {author} {\bibfnamefont {A.}~\bibnamefont {Kerelsky}}, \bibinfo
  {author} {\bibfnamefont {B.}~\bibnamefont {Kim}}, \bibinfo {author}
  {\bibfnamefont {J.}~\bibnamefont {Wang}}, \bibinfo {author} {\bibfnamefont
  {A.}~\bibnamefont {Zangiabadi}}, \bibinfo {author} {\bibfnamefont
  {C.}~\bibnamefont {Kim}}, \bibinfo {author} {\bibfnamefont {A.}~\bibnamefont
  {Abhinandan}}, \bibinfo {author} {\bibfnamefont {J.}~\bibnamefont
  {Ardelean}}, \bibinfo {author} {\bibfnamefont {M.}~\bibnamefont {Scully}},
  \bibinfo {author} {\bibfnamefont {D.}~\bibnamefont {Scullion}}, \bibinfo
  {author} {\bibfnamefont {L.}~\bibnamefont {Embon}}, \bibinfo {author}
  {\bibfnamefont {R.}~\bibnamefont {Zu}}, \bibinfo {author} {\bibfnamefont
  {E.~J.}\ \bibnamefont {Santos}}, \bibinfo {author} {\bibfnamefont
  {L.}~\bibnamefont {Balicas}}, \bibinfo {author} {\bibfnamefont
  {C.}~\bibnamefont {Marianetti}}, \bibinfo {author} {\bibfnamefont
  {K.}~\bibnamefont {Barmak}}, \bibinfo {author} {\bibfnamefont
  {X.}~\bibnamefont {Zhu}}, \bibinfo {author} {\bibfnamefont {J.}~\bibnamefont
  {Hone}}, \ and\ \bibinfo {author} {\bibfnamefont {A.~N.}\ \bibnamefont
  {Pasupathy}},\ }\href {\doibase 10.1021/acs.nanolett.9b00985} {\bibfield
  {journal} {\bibinfo  {journal} {Nano Letters}\ }\textbf {\bibinfo {volume}
  {19}},\ \bibinfo {pages} {4371} (\bibinfo {year} {2019})}\BibitemShut
  {NoStop}%
\bibitem [{\citenamefont {Liu}\ \emph {et~al.}(2023)\citenamefont {Liu},
  \citenamefont {Liu}, \citenamefont {Holtzman}, \citenamefont {Li},
  \citenamefont {Holbrook}, \citenamefont {Pack}, \citenamefont {Taniguchi},
  \citenamefont {Watanabe}, \citenamefont {Dean}, \citenamefont {Pasupathy},
  \citenamefont {Barmak}, \citenamefont {Rhodes},\ and\ \citenamefont
  {Hone}}]{Liu:2023}%
  \BibitemOpen
  \bibfield  {author} {\bibinfo {author} {\bibfnamefont {S.}~\bibnamefont
  {Liu}}, \bibinfo {author} {\bibfnamefont {Y.}~\bibnamefont {Liu}}, \bibinfo
  {author} {\bibfnamefont {L.~N.}\ \bibnamefont {Holtzman}}, \bibinfo {author}
  {\bibfnamefont {B.}~\bibnamefont {Li}}, \bibinfo {author} {\bibfnamefont
  {M.}~\bibnamefont {Holbrook}}, \bibinfo {author} {\bibfnamefont
  {J.}~\bibnamefont {Pack}}, \bibinfo {author} {\bibfnamefont {T.}~\bibnamefont
  {Taniguchi}}, \bibinfo {author} {\bibfnamefont {K.}~\bibnamefont {Watanabe}},
  \bibinfo {author} {\bibfnamefont {C.~R.}\ \bibnamefont {Dean}}, \bibinfo
  {author} {\bibfnamefont {A.}~\bibnamefont {Pasupathy}}, \bibinfo {author}
  {\bibfnamefont {K.}~\bibnamefont {Barmak}}, \bibinfo {author} {\bibfnamefont
  {D.~A.}\ \bibnamefont {Rhodes}}, \ and\ \bibinfo {author} {\bibfnamefont
  {J.}~\bibnamefont {Hone}},\ }\href@noop {} {\  (\bibinfo {year} {2023})},\
  \Eprint {http://arxiv.org/abs/2303.16290} {arXiv:2303.16290} \BibitemShut
  {NoStop}%
\bibitem [{\citenamefont {Shi}\ \emph {et~al.}(2020)\citenamefont {Shi},
  \citenamefont {Shih}, \citenamefont {Gustafsson}, \citenamefont {Rhodes},
  \citenamefont {Kim}, \citenamefont {Watanabe}, \citenamefont {Taniguchi},
  \citenamefont {Papi{\'{c}}}, \citenamefont {Hone},\ and\ \citenamefont
  {Dean}}]{Shi2020}%
  \BibitemOpen
  \bibfield  {author} {\bibinfo {author} {\bibfnamefont {Q.}~\bibnamefont
  {Shi}}, \bibinfo {author} {\bibfnamefont {E.~M.}\ \bibnamefont {Shih}},
  \bibinfo {author} {\bibfnamefont {M.~V.}\ \bibnamefont {Gustafsson}},
  \bibinfo {author} {\bibfnamefont {D.~A.}\ \bibnamefont {Rhodes}}, \bibinfo
  {author} {\bibfnamefont {B.}~\bibnamefont {Kim}}, \bibinfo {author}
  {\bibfnamefont {K.}~\bibnamefont {Watanabe}}, \bibinfo {author}
  {\bibfnamefont {T.}~\bibnamefont {Taniguchi}}, \bibinfo {author}
  {\bibfnamefont {Z.}~\bibnamefont {Papi{\'{c}}}}, \bibinfo {author}
  {\bibfnamefont {J.}~\bibnamefont {Hone}}, \ and\ \bibinfo {author}
  {\bibfnamefont {C.~R.}\ \bibnamefont {Dean}},\ }\href@noop {} {\bibfield
  {journal} {\bibinfo  {journal} {Nature Nanotechnology}\ }\textbf {\bibinfo
  {volume} {15}},\ \bibinfo {pages} {569} (\bibinfo {year} {2020})}\BibitemShut
  {NoStop}%
\bibitem [{\citenamefont {Wang}\ \emph {et~al.}(2020)\citenamefont {Wang},
  \citenamefont {Shih}, \citenamefont {Ghiotto}, \citenamefont {Xian},
  \citenamefont {Rhodes}, \citenamefont {Tan}, \citenamefont {Claassen},
  \citenamefont {Kennes}, \citenamefont {Bai}, \citenamefont {Kim},
  \citenamefont {Watanabe}, \citenamefont {Taniguchi}, \citenamefont {Zhu},
  \citenamefont {Hone}, \citenamefont {Rubio}, \citenamefont {Pasupathy},\ and\
  \citenamefont {Dean}}]{Wang2020}%
  \BibitemOpen
  \bibfield  {author} {\bibinfo {author} {\bibfnamefont {L.}~\bibnamefont
  {Wang}}, \bibinfo {author} {\bibfnamefont {E.~M.}\ \bibnamefont {Shih}},
  \bibinfo {author} {\bibfnamefont {A.}~\bibnamefont {Ghiotto}}, \bibinfo
  {author} {\bibfnamefont {L.}~\bibnamefont {Xian}}, \bibinfo {author}
  {\bibfnamefont {D.~A.}\ \bibnamefont {Rhodes}}, \bibinfo {author}
  {\bibfnamefont {C.}~\bibnamefont {Tan}}, \bibinfo {author} {\bibfnamefont
  {M.}~\bibnamefont {Claassen}}, \bibinfo {author} {\bibfnamefont {D.~M.}\
  \bibnamefont {Kennes}}, \bibinfo {author} {\bibfnamefont {Y.}~\bibnamefont
  {Bai}}, \bibinfo {author} {\bibfnamefont {B.}~\bibnamefont {Kim}}, \bibinfo
  {author} {\bibfnamefont {K.}~\bibnamefont {Watanabe}}, \bibinfo {author}
  {\bibfnamefont {T.}~\bibnamefont {Taniguchi}}, \bibinfo {author}
  {\bibfnamefont {X.}~\bibnamefont {Zhu}}, \bibinfo {author} {\bibfnamefont
  {J.}~\bibnamefont {Hone}}, \bibinfo {author} {\bibfnamefont {A.}~\bibnamefont
  {Rubio}}, \bibinfo {author} {\bibfnamefont {A.~N.}\ \bibnamefont
  {Pasupathy}}, \ and\ \bibinfo {author} {\bibfnamefont {C.~R.}\ \bibnamefont
  {Dean}},\ }\href {\doibase 10.1038/s41563-020-0708-6} {\bibfield  {journal}
  {\bibinfo  {journal} {Nature Materials}\ }\textbf {\bibinfo {volume} {19}},\
  \bibinfo {pages} {861} (\bibinfo {year} {2020})}\BibitemShut {NoStop}%
\bibitem [{\citenamefont {Shi}\ \emph {et~al.}(2022)\citenamefont {Shi},
  \citenamefont {Shih}, \citenamefont {Rhodes}, \citenamefont {Kim},
  \citenamefont {Barmak}, \citenamefont {Watanabe}, \citenamefont {Taniguchi},
  \citenamefont {Papi{\'{c}}}, \citenamefont {Abanin}, \citenamefont {Hone},\
  and\ \citenamefont {Dean}}]{Shi2022}%
  \BibitemOpen
  \bibfield  {author} {\bibinfo {author} {\bibfnamefont {Q.}~\bibnamefont
  {Shi}}, \bibinfo {author} {\bibfnamefont {E.~M.}\ \bibnamefont {Shih}},
  \bibinfo {author} {\bibfnamefont {D.}~\bibnamefont {Rhodes}}, \bibinfo
  {author} {\bibfnamefont {B.}~\bibnamefont {Kim}}, \bibinfo {author}
  {\bibfnamefont {K.}~\bibnamefont {Barmak}}, \bibinfo {author} {\bibfnamefont
  {K.}~\bibnamefont {Watanabe}}, \bibinfo {author} {\bibfnamefont
  {T.}~\bibnamefont {Taniguchi}}, \bibinfo {author} {\bibfnamefont
  {Z.}~\bibnamefont {Papi{\'{c}}}}, \bibinfo {author} {\bibfnamefont {D.~A.}\
  \bibnamefont {Abanin}}, \bibinfo {author} {\bibfnamefont {J.}~\bibnamefont
  {Hone}}, \ and\ \bibinfo {author} {\bibfnamefont {C.~R.}\ \bibnamefont
  {Dean}},\ }\href@noop {} {\bibfield  {journal} {\bibinfo  {journal} {Nature
  Nanotechnology}\ }\textbf {\bibinfo {volume} {17}},\ \bibinfo {pages} {557}
  (\bibinfo {year} {2022})}\BibitemShut {NoStop}%
\bibitem [{\citenamefont {{Das Sarma}}\ and\ \citenamefont
  {Hwang}(2013)}]{Hwang:Universal}%
  \BibitemOpen
  \bibfield  {author} {\bibinfo {author} {\bibfnamefont {S.}~\bibnamefont {{Das
  Sarma}}}\ and\ \bibinfo {author} {\bibfnamefont {E.~H.}\ \bibnamefont
  {Hwang}},\ }\href@noop {} {\bibfield  {journal} {\bibinfo  {journal} {Phys.
  Rev. B}\ }\textbf {\bibinfo {volume} {88}},\ \bibinfo {pages} {035439}
  (\bibinfo {year} {2013})}\BibitemShut {NoStop}%
\bibitem [{\citenamefont {{Das Sarma}}\ and\ \citenamefont
  {Hwang}(2014)}]{Hwang:Short}%
  \BibitemOpen
  \bibfield  {author} {\bibinfo {author} {\bibfnamefont {S.}~\bibnamefont {{Das
  Sarma}}}\ and\ \bibinfo {author} {\bibfnamefont {E.~H.}\ \bibnamefont
  {Hwang}},\ }\href@noop {} {\bibfield  {journal} {\bibinfo  {journal} {Phys.
  Rev. B}\ }\textbf {\bibinfo {volume} {89}},\ \bibinfo {pages} {121413(R)}
  (\bibinfo {year} {2014})}\BibitemShut {NoStop}%
\bibitem [{\citenamefont {Jauregui}\ \emph {et~al.}(2019)\citenamefont
  {Jauregui}, \citenamefont {Joe}, \citenamefont {Pistunova}, \citenamefont
  {Wild}, \citenamefont {High}, \citenamefont {Zhou}, \citenamefont {Scuri},
  \citenamefont {{De Greve}}, \citenamefont {Sushko}, \citenamefont {Yu},
  \citenamefont {Taniguchi}, \citenamefont {Watanabe}, \citenamefont
  {Needleman}, \citenamefont {Lukin}, \citenamefont {Park},\ and\ \citenamefont
  {Kim}}]{Jauregui:OurPaper}%
  \BibitemOpen
  \bibfield  {author} {\bibinfo {author} {\bibfnamefont {L.~A.}\ \bibnamefont
  {Jauregui}}, \bibinfo {author} {\bibfnamefont {A.~Y.}\ \bibnamefont {Joe}},
  \bibinfo {author} {\bibfnamefont {K.}~\bibnamefont {Pistunova}}, \bibinfo
  {author} {\bibfnamefont {D.~S.}\ \bibnamefont {Wild}}, \bibinfo {author}
  {\bibfnamefont {A.~A.}\ \bibnamefont {High}}, \bibinfo {author}
  {\bibfnamefont {Y.}~\bibnamefont {Zhou}}, \bibinfo {author} {\bibfnamefont
  {G.}~\bibnamefont {Scuri}}, \bibinfo {author} {\bibfnamefont
  {K.}~\bibnamefont {{De Greve}}}, \bibinfo {author} {\bibfnamefont
  {A.}~\bibnamefont {Sushko}}, \bibinfo {author} {\bibfnamefont {C.-H.}\
  \bibnamefont {Yu}}, \bibinfo {author} {\bibfnamefont {T.}~\bibnamefont
  {Taniguchi}}, \bibinfo {author} {\bibfnamefont {K.}~\bibnamefont {Watanabe}},
  \bibinfo {author} {\bibfnamefont {D.~J.}\ \bibnamefont {Needleman}}, \bibinfo
  {author} {\bibfnamefont {M.~D.}\ \bibnamefont {Lukin}}, \bibinfo {author}
  {\bibfnamefont {H.}~\bibnamefont {Park}}, \ and\ \bibinfo {author}
  {\bibfnamefont {P.}~\bibnamefont {Kim}},\ }\href {\doibase
  10.1126/science.aaw4194} {\bibfield  {journal} {\bibinfo  {journal}
  {Science}\ }\textbf {\bibinfo {volume} {366}},\ \bibinfo {pages} {870}
  (\bibinfo {year} {2019})}\BibitemShut {NoStop}%
\bibitem [{\citenamefont {Kaasbjerg}(2020)}]{Kaasbjerg2020}%
  \BibitemOpen
  \bibfield  {author} {\bibinfo {author} {\bibfnamefont {K.}~\bibnamefont
  {Kaasbjerg}},\ }\href {\doibase 10.1103/PhysRevB.101.045433} {\bibfield
  {journal} {\bibinfo  {journal} {Phys. Rev. B}\ }\textbf {\bibinfo {volume}
  {101}},\ \bibinfo {pages} {045433} (\bibinfo {year} {2020})}\BibitemShut
  {NoStop}%
\bibitem [{\citenamefont {Kaasbjerg}\ \emph {et~al.}(2012)\citenamefont
  {Kaasbjerg}, \citenamefont {Thygesen},\ and\ \citenamefont
  {Jacobsen}}]{Kaasbjerg2017}%
  \BibitemOpen
  \bibfield  {author} {\bibinfo {author} {\bibfnamefont {K.}~\bibnamefont
  {Kaasbjerg}}, \bibinfo {author} {\bibfnamefont {K.~S.}\ \bibnamefont
  {Thygesen}}, \ and\ \bibinfo {author} {\bibfnamefont {K.~W.}\ \bibnamefont
  {Jacobsen}},\ }\href {\doibase 10.1103/PhysRevB.85.115317} {\bibfield
  {journal} {\bibinfo  {journal} {Phys. Rev. B}\ }\textbf {\bibinfo {volume}
  {85}},\ \bibinfo {pages} {115317} (\bibinfo {year} {2012})}\BibitemShut
  {NoStop}%
\bibitem [{\citenamefont {Li}\ \emph {et~al.}(2018)\citenamefont {Li},
  \citenamefont {Wang}, \citenamefont {Lu}, \citenamefont {Jin}, \citenamefont
  {Chen}, \citenamefont {Meng}, \citenamefont {Lian}, \citenamefont
  {Taniguchi}, \citenamefont {Watanabe}, \citenamefont {Zhang}, \citenamefont
  {Smirnov},\ and\ \citenamefont {Shi}}]{Li2018}%
  \BibitemOpen
  \bibfield  {author} {\bibinfo {author} {\bibfnamefont {Z.}~\bibnamefont
  {Li}}, \bibinfo {author} {\bibfnamefont {T.}~\bibnamefont {Wang}}, \bibinfo
  {author} {\bibfnamefont {Z.}~\bibnamefont {Lu}}, \bibinfo {author}
  {\bibfnamefont {C.}~\bibnamefont {Jin}}, \bibinfo {author} {\bibfnamefont
  {Y.}~\bibnamefont {Chen}}, \bibinfo {author} {\bibfnamefont {Y.}~\bibnamefont
  {Meng}}, \bibinfo {author} {\bibfnamefont {Z.}~\bibnamefont {Lian}}, \bibinfo
  {author} {\bibfnamefont {T.}~\bibnamefont {Taniguchi}}, \bibinfo {author}
  {\bibfnamefont {K.}~\bibnamefont {Watanabe}}, \bibinfo {author}
  {\bibfnamefont {S.}~\bibnamefont {Zhang}}, \bibinfo {author} {\bibfnamefont
  {D.}~\bibnamefont {Smirnov}}, \ and\ \bibinfo {author} {\bibfnamefont
  {S.~F.}\ \bibnamefont {Shi}},\ }\href {\doibase 10.1038/s41467-018-05863-5}
  {\bibfield  {journal} {\bibinfo  {journal} {Nature Communications}\ }\textbf
  {\bibinfo {volume} {9}} (\bibinfo {year} {2018}),\
  10.1038/s41467-018-05863-5}\BibitemShut {NoStop}%
\bibitem [{\citenamefont {Barbone}\ \emph {et~al.}(2018)\citenamefont
  {Barbone}, \citenamefont {Montblanch}, \citenamefont {Kara}, \citenamefont
  {Palacios-Berraquero}, \citenamefont {Cadore}, \citenamefont {{De Fazio}},
  \citenamefont {Pingault}, \citenamefont {Mostaani}, \citenamefont {Li},
  \citenamefont {Chen}, \citenamefont {Watanabe}, \citenamefont {Taniguchi},
  \citenamefont {Tongay}, \citenamefont {Wang}, \citenamefont {Ferrari},\ and\
  \citenamefont {Atat{\"{u}}re}}]{Barbone2018}%
  \BibitemOpen
  \bibfield  {author} {\bibinfo {author} {\bibfnamefont {M.}~\bibnamefont
  {Barbone}}, \bibinfo {author} {\bibfnamefont {A.~R.}\ \bibnamefont
  {Montblanch}}, \bibinfo {author} {\bibfnamefont {D.~M.}\ \bibnamefont
  {Kara}}, \bibinfo {author} {\bibfnamefont {C.}~\bibnamefont
  {Palacios-Berraquero}}, \bibinfo {author} {\bibfnamefont {A.~R.}\
  \bibnamefont {Cadore}}, \bibinfo {author} {\bibfnamefont {D.}~\bibnamefont
  {{De Fazio}}}, \bibinfo {author} {\bibfnamefont {B.}~\bibnamefont
  {Pingault}}, \bibinfo {author} {\bibfnamefont {E.}~\bibnamefont {Mostaani}},
  \bibinfo {author} {\bibfnamefont {H.}~\bibnamefont {Li}}, \bibinfo {author}
  {\bibfnamefont {B.}~\bibnamefont {Chen}}, \bibinfo {author} {\bibfnamefont
  {K.}~\bibnamefont {Watanabe}}, \bibinfo {author} {\bibfnamefont
  {T.}~\bibnamefont {Taniguchi}}, \bibinfo {author} {\bibfnamefont
  {S.}~\bibnamefont {Tongay}}, \bibinfo {author} {\bibfnamefont
  {G.}~\bibnamefont {Wang}}, \bibinfo {author} {\bibfnamefont {A.~C.}\
  \bibnamefont {Ferrari}}, \ and\ \bibinfo {author} {\bibfnamefont
  {M.}~\bibnamefont {Atat{\"{u}}re}},\ }\href
  {www.nature.com/naturecommunications} {\bibfield  {journal} {\bibinfo
  {journal} {Nature Communications}\ }\textbf {\bibinfo {volume} {9}} (\bibinfo
  {year} {2018})}\BibitemShut {NoStop}%
\bibitem [{\citenamefont {Ostrovsky}\ \emph {et~al.}(2006)\citenamefont
  {Ostrovsky}, \citenamefont {Gornyi},\ and\ \citenamefont
  {Mirlin}}]{Mirlin:Electron}%
  \BibitemOpen
  \bibfield  {author} {\bibinfo {author} {\bibfnamefont {P.~M.}\ \bibnamefont
  {Ostrovsky}}, \bibinfo {author} {\bibfnamefont {I.~V.}\ \bibnamefont
  {Gornyi}}, \ and\ \bibinfo {author} {\bibfnamefont {A.~D.}\ \bibnamefont
  {Mirlin}},\ }\href@noop {} {\bibfield  {journal} {\bibinfo  {journal} {Phys.
  Rev. B}\ }\textbf {\bibinfo {volume} {74}},\ \bibinfo {pages} {235443}
  (\bibinfo {year} {2006})}\BibitemShut {NoStop}%
\bibitem [{\citenamefont {Basko}(2008)}]{Basko:Resonant}%
  \BibitemOpen
  \bibfield  {author} {\bibinfo {author} {\bibfnamefont {D.~M.}\ \bibnamefont
  {Basko}},\ }\href@noop {} {\bibfield  {journal} {\bibinfo  {journal} {Phys.
  Rev. B}\ }\textbf {\bibinfo {volume} {78}},\ \bibinfo {pages} {115432}
  (\bibinfo {year} {2008})}\BibitemShut {NoStop}%
\bibitem [{\citenamefont {Robinson}\ \emph {et~al.}(2008)\citenamefont
  {Robinson}, \citenamefont {Schomerus}, \citenamefont {{Oroszl\'any}},\ and\
  \citenamefont {Fal'ko}}]{Falko:Adsorbate}%
  \BibitemOpen
  \bibfield  {author} {\bibinfo {author} {\bibfnamefont {J.~P.}\ \bibnamefont
  {Robinson}}, \bibinfo {author} {\bibfnamefont {H.}~\bibnamefont {Schomerus}},
  \bibinfo {author} {\bibfnamefont {L.}~\bibnamefont {{Oroszl\'any}}}, \ and\
  \bibinfo {author} {\bibfnamefont {V.~I.}\ \bibnamefont {Fal'ko}},\
  }\href@noop {} {\bibfield  {journal} {\bibinfo  {journal} {Phys. Rev. Lett.}\
  }\textbf {\bibinfo {volume} {101}},\ \bibinfo {pages} {196803} (\bibinfo
  {year} {2008})}\BibitemShut {NoStop}%
\bibitem [{\citenamefont {Wehling}\ \emph {et~al.}(2010)\citenamefont
  {Wehling}, \citenamefont {Yuan}, \citenamefont {Lichtenstein}, \citenamefont
  {Geim},\ and\ \citenamefont {Katsnelson}}]{Katsnelson:Resonant}%
  \BibitemOpen
  \bibfield  {author} {\bibinfo {author} {\bibfnamefont {T.~O.}\ \bibnamefont
  {Wehling}}, \bibinfo {author} {\bibfnamefont {S.}~\bibnamefont {Yuan}},
  \bibinfo {author} {\bibfnamefont {A.~I.}\ \bibnamefont {Lichtenstein}},
  \bibinfo {author} {\bibfnamefont {A.~K.}\ \bibnamefont {Geim}}, \ and\
  \bibinfo {author} {\bibfnamefont {M.~I.}\ \bibnamefont {Katsnelson}},\
  }\href@noop {} {\bibfield  {journal} {\bibinfo  {journal} {Phys. Rev. Lett.}\
  }\textbf {\bibinfo {volume} {105}},\ \bibinfo {pages} {056802} (\bibinfo
  {year} {2010})}\BibitemShut {NoStop}%
\end{thebibliography}


\end{document}